\documentclass[%
reprint,
superscriptaddress,
amsmath,amssymb,
aps,
prl,
]{revtex4-2}
\usepackage{graphicx}
\usepackage{bm}
\usepackage{amsmath}
\newcommand{\RNum}[1]{\uppercase\expandafter{\romannumeral #1\relax}}
\begin{document}

\title{Minimum current for detachment of electrolytic bubbles}
\author{Yixin Zhang}
  \affiliation{Physics of Fluids Group, Max Planck Center Twente for Complex Fluid Dynamics and J. M. Burgers Centre for Fluid Dynamics, University of Twente, P.O. Box 217, 7500 AE Enschede, The Netherlands}
\author{Detlef Lohse}
\email{d.lohse@utwente.nl}
  \affiliation{Physics of Fluids Group, Max Planck Center Twente for Complex Fluid Dynamics and J. M. Burgers Centre for Fluid Dynamics, University of Twente, P.O. Box 217, 7500 AE Enschede, The Netherlands}
  \affiliation{Max Planck Institute for Dynamics and Self-Organization, 37077 Göttingen, Germany}
\begin{abstract}
The efficiency of water electrolysis is significantly impacted by the generation of micro- and nanobubbles on the electrodes. Here molecular dynamics simulations are used to investigate the dynamics of single electrolytic nanobubbles on nanoelectrodes. The simulations reveal that, depending on the value of current, nucleated nanobubbles either grow to an equilibrium state or grow unlimitedly and then detach. To account for these findings, the stability theory for surface nanobubbles is generalized by incorporating the electrolytic gas influx at the nanobubble's contact line and adopting a real gas law, leading to accurate predictions for the numerically observed transient growth and stationary states of the nanobubbles. With this theory, the minimum current for bubble detachment can also be analytically derived. In the detachment regime, the radius of the nanobubble first increases as $R\propto t^{1/2}$ and then as $R\propto t^{1/3}$, up to bubble detachment. 
\end{abstract} 
\maketitle 
 \newpage
Hydrogen produced by water electrolysis using gas-evolving electrodes from renewable electricity is essential for achieving carbon neutrality and a sustainable future. However, micro- and nanobubbles formed at an electrode can result in undesired blockage of the electrode and thus decrease the energy transformation efficiency\,\citep{vogt2005bubble,angulo2020influence,zhao2019gas}. Addressing this problem requires a deeper understanding of the dynamics of individual nanobubbles on nanoelectrodes, which has been investigated using experiments\,\citep{luo2013electrogeneration,liu2017dynamic,edwards2019voltammetric} and molecular dynamics simulations\,\citep{perez2019mechanisms,maheshwari2020nucleation,ma2021dynamic}. However, a fundamental theory for the statics and dynamics of single electrolytic nanobubbles is still lacking. Electrolytic nanobubbles belong to the family of surface nanobubbles. Since their discovery in the 1990s, surface nanobubbles have triggered the fascination of scientists\,\citep{lohse2015surface} due to their mysterious properties such as their long lifetime\,\citep{weijs2013surface,lou2000nanobubbles} and their small contact angles\,\citep{zhang2006physical}. Significant progress has been made (see the reviews\,\citep{lohse2015surface,tan2021stability}) in developing new methods for generating and detecting nanobubbles, as well as clarifying their stability mechanism by an analytical model developed by Lohse and Zhang\,\citep{lohse2015pinning} (and extensions thereof), which suggests that a stable balance between the Laplace pressure-driven gas outflux and the local oversaturation-driven gas influx is made possible by contact line pinning. 

This Letter aims to understand the dynamics of single electrolytic nanobubbles through molecular dynamics simulations and analytical theories. The Lohse-Zhang model is generalized to account for the gas produced at the contact line. The new model can quantitatively explain the conducted molecular simulations without any free parameters. This generalization thus creates a unified theoretical framework that can predict not only the equilibrium states of stable electrolytic nanobubbles (such as contact angles and the time required to reach equilibrium), but also the unbounded growth of unstable electrolytic nanobubbles, which eventually detach. 
\begin{figure}[t]
\includegraphics [width=\linewidth]{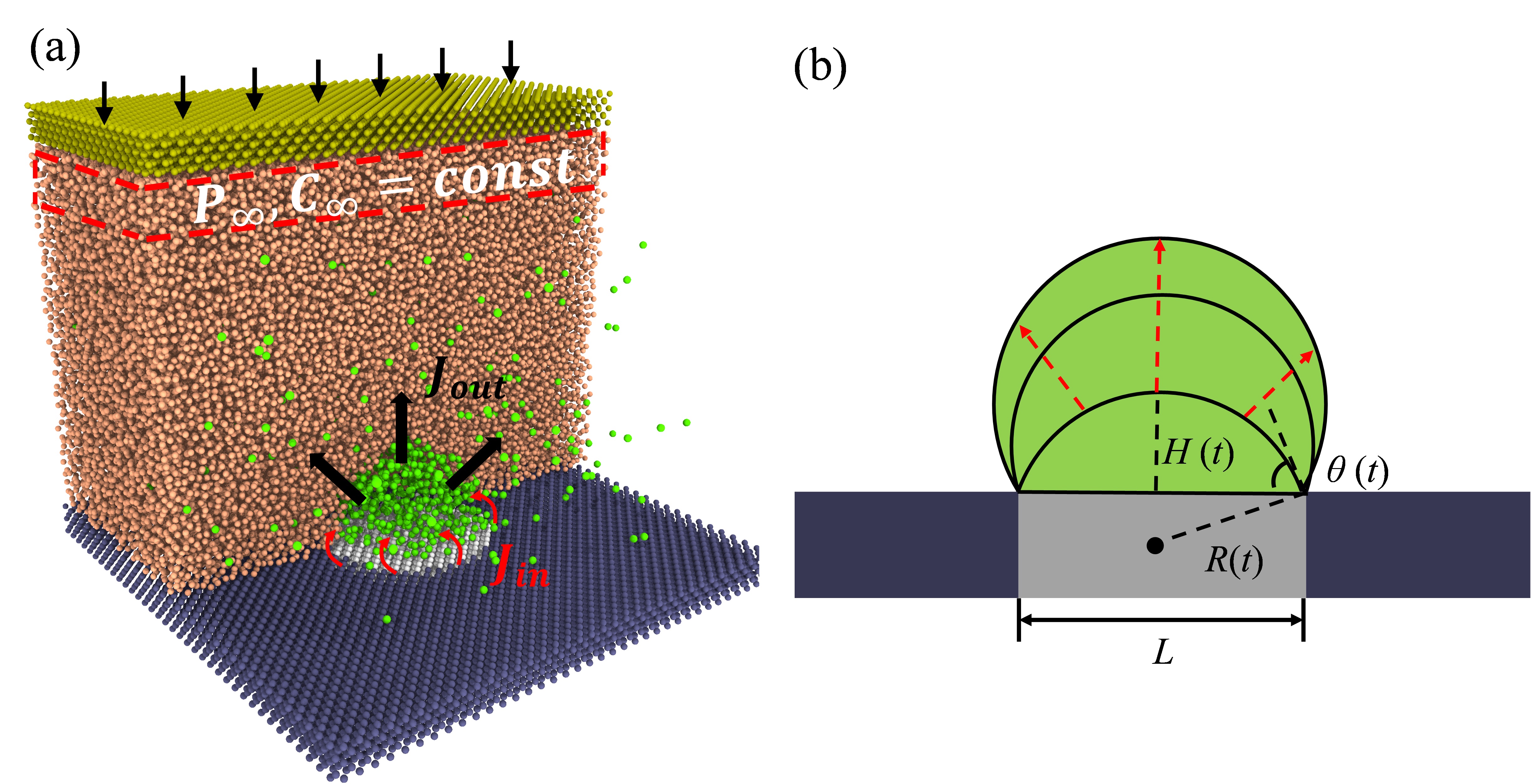}
\caption{\label{fig1} (a) A snapshot of the generated electrolytic nanobubble on the nanoelectrode in MD simulations. The simulated domain has been sliced to observe the bubble. The system's condition is maintained at $T=300$ K, $P_{\infty}=10$ atm, and $C_{\infty}=c_s$. The nanoelectrode has a diameter $L=5.76$ nm and is made hydrophobic to water, while the surrounding solid is hydrophilic. $J_{in}$ represents the gas influx to the bubble produced at the contact line, while $J_{out}$ represents the diffusive outflux through the bubble interface. (b) Sketch of a growing bubble with pinning length $L$, contact angle $\theta\left(t\right)$, radius of curvature $R\left(t\right)$ and height $H\left(t\right)$.}
\end{figure}

\emph{MD Simulations}.\,--- Molecular dynamics (MD) simulations are used as virtual experiments to simulate the generation of electrolytic nanobubbles on nanoelectrodes. We adopt the open-source code LAMMPS\,\cite{plimpton1995fast}. All used parameters, simulation strategies and validation are provided in detail in the Supplemental Materials (SM). As shown in Fig.~\ref{fig1}(a), the minimal molecular system consists of water molecules (represented in orange), gas atoms (in green), atoms of the electrode (in white), atoms of the solid base (in blue) and atoms of the `piston' plate (in bronze). The water is modeled by the mW water potential\,\citep{molinero2009water}, implying a surface tension $\gamma=66$ mN/m. The gas modeled by the standard 12-6 Lennard-Jones (LJ) potential has a density $\rho_{\infty}=11.47$ kg/m\textsuperscript{3} at 10 atm and 300 K. The gas-water interaction is tuned to obtain a gas solubility $c_s=0.54$ kg/m\textsuperscript{3} and a mass diffusivity $D=4.1\times 10^{-9}$ m\textsuperscript{2}/s. The electrochemical reaction that transforms water molecules into gas atoms is modeled in a simple way like previous MD studies\,\citep{perez2019mechanisms,maheshwari2020nucleation,ma2021dynamic} that right above the electrodes, three layers of water molecules can turn into gas atoms conducted at a fixed rate, leading to a constant gas influx $J_{in}$, i.e., a constant current $i_{in}$ (assuming the production of each gas atom needs a specific number of electrons). As sketched in Fig.~\ref{fig1}(a) and (b), after nucleation the bubble grows until the balance between the Laplace pressure-driven diffusive outfluxes ($J_{out}$) and the reaction-driven gas influxes ($J_{in}$) is achieved (which may not happen). The evolution of the bubble's contact angle $\theta(t)$, radius of curvature $R(t)$, and height $H(t)$ will be of particular interest in this study.

We find that for very small gas influxes, no bubbles can nucleate due to the low level of oversaturation around the electrodes. The gas concentration in steady states close to the electrode may be estimated by $c_L=c_s+J_{in}/\left(2DL\right)$\,\citep{saito1968theoretical} so that a critical concentration for the bubble to nucleate requires a critical gas influx. In our simulations, for $J_{in}> 1\times 10^{-15}$ kg/s (corresponding to about 100 times oversaturation), nucleation always happens in agreement with the critical oversaturation found in experiments\,\citep{chen2014electrochemical}. The initial contact angle $\theta_i$ for the pinned bubble may be estimated by assuming that the bubble is composed of one layer of gas atoms with height $0.375$ nm and using the geometric relation $H=L\left(1-\cos\theta\right)/\left(2\sin\theta\right)\approx L/\left(4\theta\right)$, resulting in  $\theta_i=15^{\circ}$ very similar to the $\theta_i=20^{\circ}$ determined by the voltammetric method in experiments\,\citep{edwards2019voltammetric}. Such a small $\theta_i$ makes it possible to study the evolution of contact angles of nanobubbles.
\begin{figure}[h!]
\includegraphics [width=\linewidth]{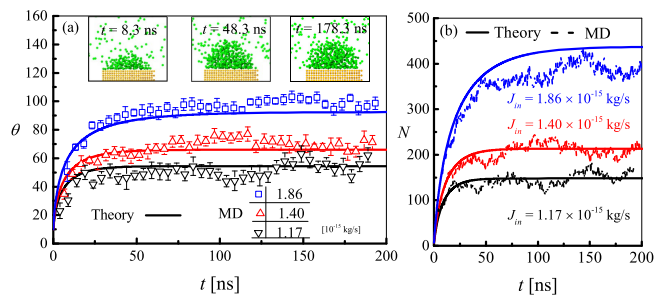}
\caption{\label{fig2} (a) Growth of contact angles to their equilibrium states for three different gas influxes $J_{in}$ as given in the legend. The symbols represent MD results and solid lines are obtained by solving Eq.~\eqref{eq:theta_evol}. The MD snapshots show the evolution of the simulated nanobubble for the case $J_{in}=1.86\times 10^{-15}$ kg/s at three different times. (b) Growth of the number of gas atoms in the bubble $N(t)$ to the equilibrium state for the same three different gas influxes.}
\end{figure}

\begin{center}
\begin{figure}[t!]
\includegraphics [width=\linewidth]{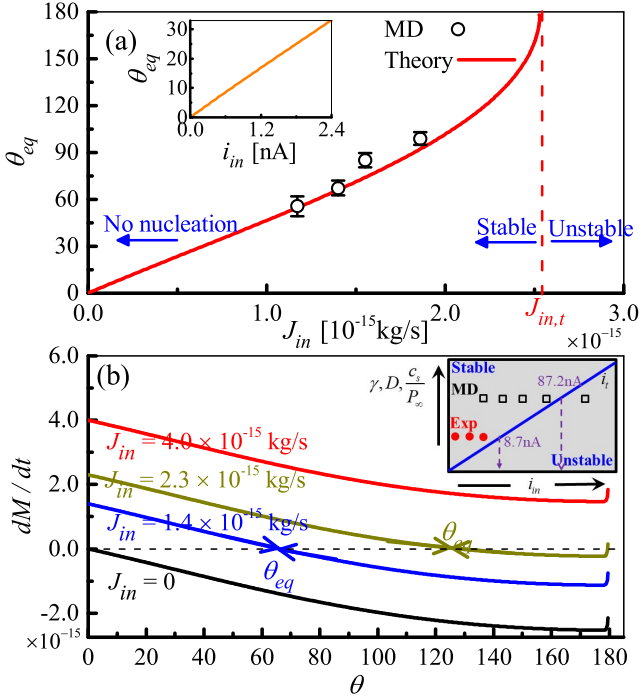}
\caption{\label{fig3} (a) The relation between equilibrium contact angles $\theta_{eq}$ and gas influxe $J_{in}$. The symbols represent the results obtained from MD simulations. The solid line is the theoretical prediction, i.e., Eq.~\eqref{eq:angle_equi_exact}. $J_t$ is the calculated threshold gas influx, differentiating between stable and unstable nanobubbles. The inset shows predictions for $\theta_{eq}$ of electrolytic bubbles in experiments\,\cite{liu2017dynamic} using the measured currents $i_{in}$, indicating that nanobubbles in experiments are indeed stable. (b) Dependence of the mass change rate on the contact angle for four different gas influxes. The inset shows a sketch of the phase diagram for stable and unstable nanobubbles observed in MD and experiments\,\cite{liu2017dynamic}. The dividing line is given by the threshold current Eq.~\eqref{eq:it}.}
\end{figure}
\end{center}

Indeed, for example, in the case of $J_{in}=1.86\times 10^{-15}$ kg/s, its MD snapshots in Fig.~\ref{fig2}(a) (also see the video in SM) demonstrates the transient growth of the pinned nanobubble to its stationary state. The instantaneous contact angle $\theta(t)$ of the nanobubble is then obtained by fitting the instantaneous liquid-gas interface with a spherical cap (see SM-\RNum{4}) and is shown in Fig.~\ref{fig2}(a) for three different gas influxes $J_{in}$. It can be seen that for all cases $\theta(t)$ increases with time initially but reaches its steady state eventually. The equilibrated state can also be examined by tracking the number of gas atoms $N(t)$ in the bubble shown in Fig.~\ref{fig2}(b), which is obtained by counting the gas atoms below the instantaneous liquid-gas interface (see SM-\RNum{4}). After verifying that the nanobubbles are at equilibrium, the equilibrium contact angles $\theta_{eq}$ for different $J_{in}$ are then obtained by averaging the data in the last $20$ ns of each simulation and are shown in Fig.~\ref{fig3}(a). For even larger gas influxes, e.g., $J_{in}=4.5\times 10^{-15}$ kg/s, the nucleated nanobubble is unstable and it becomes so large that it comes into contact with its periodic images in simulations, see Fig.~S4 in SM.
 
\emph{Generalized Lohse-Zhang equation}. --- To predict the transient growth and stationary states of electrolytic nanobubbles from MD simulations, we propose the following model. The longevity of surface nanobubbles is due to the contact line pinning and the local gas oversaturation as described by the Lohse-Zhang equation\,\citep{lohse2015pinning}, derived in analogy to the problem of droplet evaporation\,\citep{popov2005evaporative}. After nucleation, the single electrolytic nanobubble forming on the nanoelectrode gets pinned at the edge of the electrode due to the material heterogeneity between the electrode and the surrounding solid. The pinned nanobubble leads to the blockage of the water access to the electrode, leaving only the region of the bubble's contact line to produce gas. The produced gas $J_{in}$ may enter directly into the bubble due to energy minimization. We reflect this in a generalization of the Lohse-Zhang equation in order to account for the gas production:
\begin{equation}\label{eq:mass_rate}
\frac{dM}{dt}=-\frac{\pi }{2}LD{{c}_{s}}\left( \frac{4\gamma }{L{{P}_{\infty}}}\sin \theta -\zeta  \right)f\left( \theta  \right)+{{J}_{in}},\,\,\,\mbox{with}
\end{equation}
\begin{equation}
f\left( \theta  \right)=\frac{\sin \theta }{1+\cos \theta }+4\int_{0}^{\infty }{\frac{1+\cosh 2\theta \xi }{\sinh 2\pi \xi }}\tanh \left[ \left( \pi -\theta  \right)\xi  \right]d\xi.
\end{equation}
Here $J_{in}$ is the reaction-controlled gas influx at the contact line and is the new term as compared to the original Lohse-Zhang equation. $M$ is the mass of the bubble. $\zeta=c_{\infty}/c_s-1$ is the gas oversaturation and we choose the saturated situation $\zeta=0$ as in real experiments and our simulations. The first term in the right hand side of Eq.~\eqref{eq:mass_rate} is the diffusive outflux ($J_{out}$) through the bubble surface. Its relation with $\theta$ is shown in Fig.~\ref{fig3}(b) (black solid line for $J_{in}=0$).

The transient dynamics of the nanobubbles depends on the equation of state for gas atoms in the nanobubbles. For nanobubbles whose radii are as small as a few nanometers, the inside pressure $P_R$ can be tens of millions Pascals (in our case, the maximum $P_R\approx 47$ MPa for $L=5.76$ nm) so that the ideal gas law breaks down (see Fig.~S5 in SM). To account for the volume occupied by gas atoms and interatomic potentials, we adopt the Van der Waals equation as a real gas law, $N=PV/\left(P b+k_BT\right)$\,\citep{silbey2022physical}, where $N$ is the number of atoms, $V$ is the total volume, $b=4\pi\sigma_{eff}^3/3$ is the volume per atom with an effective atomic radius $\sigma_{eff}=0.2$ nm, and $k_B$ is the Boltzmann constant. Note that the modification to $P$ by the interatomic interaction is included by the effective atomic radius. The bubble density thus is
\begin{equation}\label{eq:gas_law}
{{\rho }_{R}}={{\rho }_{\infty }}\left( 1+\frac{2\gamma }{R{{P}_{\infty }}} \right)\frac{{{P}_{\infty }}{b}+{{k}_{B}}T}{\left( 1+\frac{2\gamma }{R{{P}_{\infty }}} \right){{P}_{\infty }}{b}+{{k}_{B}}T}.
\end{equation}
Considering that $R=L/\left(2\sin\theta\right)$ and that the bubble's volume is $V_b={\pi{L}^{3} }({{{\cos }^{3}}\theta -3\cos \theta +2})/({24{{\sin }^{3}}\theta })$, we get
\begin{equation}\label{eq:mass_mass}
M=\frac{{{\rho }_{\infty }}\left( {{P}_{\infty }}{b}+{{k}_{B}}T \right)\left( {\pi }{{L}^{3}}\frac{{{\cos }^{3}}\theta -3\cos \theta +2}{24{{\sin }^{3}}\theta } \right)}{{{P}_{\infty }}{b}+\frac{{{k}_{B}}T}{1+\frac{4\gamma }{L{{P}_{\infty }}}\sin \theta }}.
\end{equation}
Substituting Eq.~\eqref{eq:mass_mass} into Eq.~\eqref{eq:mass_rate} leads to an ODE for $\theta(t)$
\begin{align}\label{eq:theta_evol}
  & \frac{d\theta }{dt}=\frac{-\frac{\pi }{2}LD{{c}_{s}}\left( \frac{{4\gamma}}{LP_{\infty}}\sin \theta -\zeta  \right)f\left( \theta  \right)+{{J}_{in}}}{\frac{\frac{\pi }{8}{{L}^{3}}{{\rho }_{\infty }}\left( {{P}_{\infty }}{b}+{{k}_{B}}T \right)}{\left( 1+\frac{4\gamma }{L{{P}_{\infty }}}\sin \theta  \right){{P}_{\infty }}{b}+{{k}_{B}}T}\left( {{T}_{1}}-{{T}_{2}} \right)}, \,\,\,\mbox{with} \\ 
 & {{T}_{1}}=\frac{1+\frac{4\gamma }{L{{P}_{\infty }}}\sin \theta }{{{\left( 1+\cos \theta  \right)}^{2}}}+\frac{4\gamma }{L{{P}_{\infty }}}\frac{{{\cos }^{3}}\theta -3\cos \theta +2}{3{{\sin }^{3}}\theta }\cos \theta,  \\ 
 & {{T}_{2}}=\frac{\frac{4\gamma {b}}{L}\cos \theta \left( 1+\frac{4\gamma }{L{{P}_{\infty }}}\sin \theta  \right)\left( \frac{{{\cos }^{3}}\theta -3\cos \theta +2}{3{{\sin }^{3}}\theta } \right)}{\left( 1+\frac{4\gamma }{L{{P}_{\infty }}}\sin \theta  \right){{P}_{\infty }}b+{{k}_{B}}T}.
\end{align}

 The dynamical evolution of $\theta(t)$ and $N(t)$ are easily obtained by numerical solutions to Eq.~\eqref{eq:theta_evol}. The results agree well with the results from the previous MD simulations (see solid lines in Fig.~\ref{fig2}(a) and Fig.~\ref{fig2}(b)). The time $t_{eq}$ it takes for the nucleated nanobubble to reach equilibrium increases sharply with the value of the gas influx $J_{in}$ (see Fig.~S6 in SM) and the size of electrodes $L$ ($t_{eq}\propto L^n$, $n=2\sim3$, see Fig.~S7 in SM) and it becomes inaccessible in MD simulations due to the restrictive computational costs. This highlights the importance of the newly developed model to predict the evolution of electrolytic nanobubbles.

\emph{Threshold gas influx $J_{in,t}$ for stable and unstable nanobubbles}. --- The competition between the influx $J_{in}$ and the outflux $J_{out}$ leads to the possible existence of stable surface nanobubbles. As illustrated in Fig.~\ref{fig3}(b), adding a small value of $J_{in}=1.4\times 10^{-15}$ kg/s to $dM/dt$ allows the blue line to cross the horizontal dash line of zero-mass change rate and the intersection point with the equilibrium angel $\theta_{eq}$ is indeed stable, as any deviations from $\theta_{eq}$ will lead to a mass flux bringing the angle back to $\theta_{eq}$ (see the arrows). Note that a further increase of the influx to $J_{in}=2.3\times 10^{-15}$ kg/s leads to a bifurcation and two intersection points where $dM/dt=0$ (see the bronze line), but only the first point is stable. By putting $dM/dt=0$ in Eq.~\eqref{eq:mass_rate}, we obtain an implicit expression for the equilibrium contact angle,
\begin{equation} \label{eq:angle_equi_exact}
\sin \theta_{eq} f\left( \theta_{eq}  \right)=\frac{{{P}_{\infty}}}{2\pi D{{c}_{s}}\gamma }{{J}_{in}}~,
\end{equation}
whose solutions for different $J_{in}$ compare excellently with the equilibrium contact angles measured from MD simulations as shown in Fig.~\ref{fig3}(a). To obtain an explicit equation for small contact angles, Eq.~\eqref{eq:angle_equi_exact} can be simplified to $\theta_{eq}\approx {{J}_{in}}{{{P}_{\infty}}}/\left({8 D{{c}_{s}}\gamma }\right)$. Practically this approximation works well for $\theta \le \pi/2 $ (see Fig.~S8 in SM). We also remark that for small gas influx $J_{in}$ there is no bubble nucleation on the nanoelectrode but the exact value of the critical $J_{in}$ for bubble nucleation is not explored in this work.

For large gas influxes, nanobubbles cannot be stable. This can be seen from Fig.~\ref{fig3}(b) as adding a very large influx $J_{in}=4.0\times 10^{-15}$ kg/s (see the red line) makes $dM/dt$ always positive (above the dash line of zero-mass rate) for any angles so that no stable angles can exist. We also refer to Fig.~\ref{fig3}(a) where no equilibrium angles can be obtained for $J_{in}>J_{in,t}$. The threshold influx $J_{in,t}$ differentiating stable nanobubbles from unstable nanobubbles is obtained as
\begin{equation}\label{eq:jt}
J_{in,t}=\frac{2\pi \gamma D c_s }{P_{\infty}}\mathrm{max}\left[\sin\theta f\left(\theta\right)\right]\approx \frac{5.6\pi \gamma D c_s }{P_{\infty}},
\end{equation}
where the numerical value of $\mathrm{max}\left[\sin\theta f\left(\theta\right)\right]\approx 2.8$ has been used.

By multiplying $J_{in,t}$ with $nF/M_g$ ($n$ is the number of electrons transferred for each gas atom, $F$ is the Faraday constant, and $M_g$ is the gas molar mass), the minimum (threshold) electric current for unstable nanobubbles is 
\begin{equation}\label{eq:it}
i_{t}\approx\frac{5.6\pi \gamma D c_s nF }{P_{\infty}M_g}.
\end{equation}
One can see from Eq.~\eqref{eq:it} and the phase diagram in the inset of Fig.~\ref{fig3}(b) that electrolytic nanobubbles are more likely to be unstable for electrolytes with lower surface tension and gas with lower mass diffusivity and solubility, since the corresponding threshold current (influx) is smaller.
\begin{figure}[h]
\includegraphics [width=\linewidth]{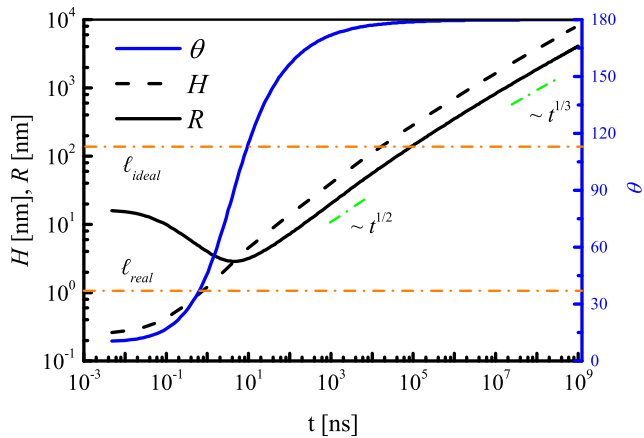}
\caption{\label{fig4} Unbounded growth of nanobubbles when the gas influx $J_{in}=5.2\times 10^{-15}$ kg/s is beyond the threshold influx $J_{in,t}=2.53\times 10^{-15}$ kg/s. For $\ell_{real}\ll R\ll \ell_{ideal}$, the scaling $R\sim t^{1/2}$ is found while for $R\gg \ell_{ideal}$, $R\sim t^{1/3}$ .}
\end{figure}

Beyond this threshold influx ($J_{in,t}=2.53\times 10^{-15}$ kg/s, corresponding to $i_{t}=87.2$ nA for our simulations assuming $n=1$), the nucleated nanobubble becomes unstable and can grow without bounds up to detachment from the electrode by buoyancy. Through the numerical solutions to Eq.~\eqref{eq:theta_evol} and using $R=L/\left(2\sin\theta\right)$ and $H=L\left(1-\cos\theta\right)/\left(2\sin\theta\right)$, Fig.~\ref{fig4} shows the evolution of the bubble's $\theta$, $R$ and $H$ for the case $J_{in}=5.2\times 10^{-15}$ kg/s. It can be seen that after a specific time, the bubble grows with a contact angle approaching $180^{\circ}$, a constant net influx ($J_{in}+J_{out}$), and a fully spherical shape (not only a cap) whose volume is $\frac{4}{3}\pi R^3$. Therefore the growth of the bubble is governed by $d\left(\frac{4}{3}\pi R^3\rho_R\right)/dt=const$. The gas law Eq.~\eqref{eq:gas_law} determines two length scales, $\ell_{ideal}=\frac{2\gamma}{P_{\infty}}$ and $\ell_{real}=\frac{2\gamma b}{k_BT}$. For a large bubble $R\gg \ell_{idel}$, $\rho_R$ is constant so that we have the usual `reaction-controlled' scaling\,\citep{van2017electrolysis} for the bubble growth
\begin{equation}
R\sim t^{1/3}, \quad \mbox{for } \quad R\gg \ell_{ideal},
\end{equation}
However, for nanobubbles whose sizes can lay between $\ell_{real}\ll R\ll \ell_{idel}$, the gas density is $\rho_R\sim R^{-1}$ so that one can obtain   
\begin{equation}
R\sim t^{1/2},\quad \mbox{for } \quad\ell_{real}\ll R\ll \ell_{ideal}.
\end{equation}
Interestingly but incidentally, this new scaling $R\sim t^{1/2}$ is the same as the one in the `diffusion-controlled' growth of bubbles\,\citep{van2017electrolysis}. But it is obvious that the nanobubbles studied here are always `reaction-controlled'.
For $R\ll \ell_{real}$, the gas density is constant again but the pinned bubble is only a spherical cap.
As the life cycle of an unstable nanobubble after birth can involve both regimes $\ell_{real}\ll R\ll \ell_{idel}$ and $R\gg \ell_{idel}$, the growth of the nanobubble experiences a transition from $R\sim t^{1/2}$ to $R\sim t^{1/3}$, as shown in Fig.~\ref{fig4}. In this work, $\ell_{ideal}=130$ nm and $\ell_{real}=1.0$ nm (for $P_{\infty}=10$ atm), so that the applicable region of the scaling $R\sim t^{1/2}$ is narrow. However, if $P_{\infty}=1$ atm, $\ell_{ideal}=1300$ nm, leading to a much wider region to observe $R\sim t^{1/2}$. 

\emph{Connections with experiments}. --- The work of White's group has focused on generating single surface nanobubbles on nanoelectrodes with radii $<50$ nm\,\citep{luo2013electrogeneration,liu2017dynamic,edwards2019voltammetric}. Unfortunately, direct visual observation of the evolving shape of nanobubbles is not possible due to technical and principal limitations. Based on the experimental data, the (constant) residual currents are $0.2\sim 2.4$ nA for different conditions\,\citep{liu2017dynamic}, which translates into hydrogen influxes ranging from $2.08\times 10^{-18}$ kg/s to $24.96\times 10^{-18}$ kg/s. Using the water surface tension 72 mN/m, hydrogen solubility $1.6\times 10^{-3}$ kg/m\textsuperscript{3}, and mass diffusivity $4.5\times 10^{-9}$ m\textsuperscript{2}/s\,\citep{cussler2009diffusion}, the calculated contact angles based on our newly developed theory are between $3$ and $33$ degrees (also see the inset of Fig.~\ref{fig3}(a)), indicating that the generated nanobubbles are indeed stable. Note that the threshold current in experiments is calculated to be $8.7$ nA and the experimental and MD results can be put together in the phase diagram in the inset of Fig.~\ref{fig3}(b). The predicted small contact angles also agree with another experiment of electrolytic nanobubbles on highly ordered pyrolytic graphites\,\citep{yang2009electrolytically}. However, it has been shown that charges\,\citep{ma2022ion} and contamination like surfactants\,\cite{lohse2015surface} may influence the behaviors of nanobubbles, which remains to be investigated in the future.

In summary, the stability mechanism of electrolytic nanobubbles on nanoelectrodes has been explained by molecular simulations and the generalized Lohse-Zhang equation. The evolution of nanobubbles in molecular simulations can be nicely predicted by the new theory. We show that a minimum current (gas influx) is needed for nucleated nanobubbles to grow boundlessly so that they can detach from electrodes. In terms of stable nanobubbles, the relation between equilibrium contact angles and gas influxes is derived. We hope that this numerical and theoretical work will aid to develop improved methods to enhance bubble detachment and thus increase the efficiency of water electrolysis.

We are grateful for the discussions with Andrea Prosperetti and acknowledge the financial support by NWO under the project of ECCM KICKstart DE-NL. 

\begin{thebibliography}{28}%
\makeatletter
\providecommand \@ifxundefined [1]{%
 \@ifx{#1\undefined}
}%
\providecommand \@ifnum [1]{%
 \ifnum #1\expandafter \@firstoftwo
 \else \expandafter \@secondoftwo
 \fi
}%
\providecommand \@ifx [1]{%
 \ifx #1\expandafter \@firstoftwo
 \else \expandafter \@secondoftwo
 \fi
}%
\providecommand \natexlab [1]{#1}%
\providecommand \enquote  [1]{``#1''}%
\providecommand \bibnamefont  [1]{#1}%
\providecommand \bibfnamefont [1]{#1}%
\providecommand \citenamefont [1]{#1}%
\providecommand \href@noop [0]{\@secondoftwo}%
\providecommand \href [0]{\begingroup \@sanitize@url \@href}%
\providecommand \@href[1]{\@@startlink{#1}\@@href}%
\providecommand \@@href[1]{\endgroup#1\@@endlink}%
\providecommand \@sanitize@url [0]{\catcode `\\12\catcode `\$12\catcode
  `\&12\catcode `\#12\catcode `\^12\catcode `\_12\catcode `\%12\relax}%
\providecommand \@@startlink[1]{}%
\providecommand \@@endlink[0]{}%
\providecommand \url  [0]{\begingroup\@sanitize@url \@url }%
\providecommand \@url [1]{\endgroup\@href {#1}{\urlprefix }}%
\providecommand \urlprefix  [0]{URL }%
\providecommand \Eprint [0]{\href }%
\providecommand \doibase [0]{https://doi.org/}%
\providecommand \selectlanguage [0]{\@gobble}%
\providecommand \bibinfo  [0]{\@secondoftwo}%
\providecommand \bibfield  [0]{\@secondoftwo}%
\providecommand \translation [1]{[#1]}%
\providecommand \BibitemOpen [0]{}%
\providecommand \bibitemStop [0]{}%
\providecommand \bibitemNoStop [0]{.\EOS\space}%
\providecommand \EOS [0]{\spacefactor3000\relax}%
\providecommand \BibitemShut  [1]{\csname bibitem#1\endcsname}%
\let\auto@bib@innerbib\@empty
\bibitem [{\citenamefont {Vogt}\ and\ \citenamefont
  {Balzer}(2005)}]{vogt2005bubble}%
  \BibitemOpen
  \bibfield  {author} {\bibinfo {author} {\bibfnamefont {H.}~\bibnamefont
  {Vogt}}\ and\ \bibinfo {author} {\bibfnamefont {R.}~\bibnamefont {Balzer}},\
  }\bibfield  {title} {\bibinfo {title} {The bubble coverage of gas-evolving
  electrodes in stagnant electrolytes},\ }\href@noop {} {\bibfield  {journal}
  {\bibinfo  {journal} {Electrochim. Acta}\ }\textbf {\bibinfo {volume} {50}},\
  \bibinfo {pages} {2073} (\bibinfo {year} {2005})}\BibitemShut {NoStop}%
\bibitem [{\citenamefont {Angulo}\ \emph {et~al.}(2020)\citenamefont {Angulo},
  \citenamefont {van~der Linde}, \citenamefont {Gardeniers}, \citenamefont
  {Modestino},\ and\ \citenamefont {Rivas}}]{angulo2020influence}%
  \BibitemOpen
  \bibfield  {author} {\bibinfo {author} {\bibfnamefont {A.}~\bibnamefont
  {Angulo}}, \bibinfo {author} {\bibfnamefont {P.}~\bibnamefont {van~der
  Linde}}, \bibinfo {author} {\bibfnamefont {H.}~\bibnamefont {Gardeniers}},
  \bibinfo {author} {\bibfnamefont {M.}~\bibnamefont {Modestino}},\ and\
  \bibinfo {author} {\bibfnamefont {D.~F.}\ \bibnamefont {Rivas}},\ }\bibfield
  {title} {\bibinfo {title} {Influence of bubbles on the energy conversion
  efficiency of electrochemical reactors},\ }\href@noop {} {\bibfield
  {journal} {\bibinfo  {journal} {Joule}\ }\textbf {\bibinfo {volume} {4}},\
  \bibinfo {pages} {555} (\bibinfo {year} {2020})}\BibitemShut {NoStop}%
\bibitem [{\citenamefont {Zhao}\ \emph {et~al.}(2019)\citenamefont {Zhao},
  \citenamefont {Ren},\ and\ \citenamefont {Luo}}]{zhao2019gas}%
  \BibitemOpen
  \bibfield  {author} {\bibinfo {author} {\bibfnamefont {X.}~\bibnamefont
  {Zhao}}, \bibinfo {author} {\bibfnamefont {H.}~\bibnamefont {Ren}},\ and\
  \bibinfo {author} {\bibfnamefont {L.}~\bibnamefont {Luo}},\ }\bibfield
  {title} {\bibinfo {title} {Gas bubbles in electrochemical gas evolution
  reactions},\ }\href@noop {} {\bibfield  {journal} {\bibinfo  {journal}
  {Langmuir}\ }\textbf {\bibinfo {volume} {35}},\ \bibinfo {pages} {5392}
  (\bibinfo {year} {2019})}\BibitemShut {NoStop}%
\bibitem [{\citenamefont {Luo}\ and\ \citenamefont
  {White}(2013)}]{luo2013electrogeneration}%
  \BibitemOpen
  \bibfield  {author} {\bibinfo {author} {\bibfnamefont {L.}~\bibnamefont
  {Luo}}\ and\ \bibinfo {author} {\bibfnamefont {H.~S.}\ \bibnamefont
  {White}},\ }\bibfield  {title} {\bibinfo {title} {Electrogeneration of single
  nanobubbles at sub-50-nm-radius platinum nanodisk electrodes},\ }\href@noop
  {} {\bibfield  {journal} {\bibinfo  {journal} {Langmuir}\ }\textbf {\bibinfo
  {volume} {29}},\ \bibinfo {pages} {11169} (\bibinfo {year}
  {2013})}\BibitemShut {NoStop}%
\bibitem [{\citenamefont {Liu}\ \emph {et~al.}(2017)\citenamefont {Liu},
  \citenamefont {Edwards}, \citenamefont {German}, \citenamefont {Chen},\ and\
  \citenamefont {White}}]{liu2017dynamic}%
  \BibitemOpen
  \bibfield  {author} {\bibinfo {author} {\bibfnamefont {Y.}~\bibnamefont
  {Liu}}, \bibinfo {author} {\bibfnamefont {M.~A.}\ \bibnamefont {Edwards}},
  \bibinfo {author} {\bibfnamefont {S.~R.}\ \bibnamefont {German}}, \bibinfo
  {author} {\bibfnamefont {Q.}~\bibnamefont {Chen}},\ and\ \bibinfo {author}
  {\bibfnamefont {H.~S.}\ \bibnamefont {White}},\ }\bibfield  {title} {\bibinfo
  {title} {The dynamic steady state of an electrochemically generated
  nanobubble},\ }\href@noop {} {\bibfield  {journal} {\bibinfo  {journal}
  {Langmuir}\ }\textbf {\bibinfo {volume} {33}},\ \bibinfo {pages} {1845}
  (\bibinfo {year} {2017})}\BibitemShut {NoStop}%
\bibitem [{\citenamefont {Edwards}\ \emph {et~al.}(2019)\citenamefont
  {Edwards}, \citenamefont {White},\ and\ \citenamefont
  {Ren}}]{edwards2019voltammetric}%
  \BibitemOpen
  \bibfield  {author} {\bibinfo {author} {\bibfnamefont {M.~A.}\ \bibnamefont
  {Edwards}}, \bibinfo {author} {\bibfnamefont {H.~S.}\ \bibnamefont {White}},\
  and\ \bibinfo {author} {\bibfnamefont {H.}~\bibnamefont {Ren}},\ }\bibfield
  {title} {\bibinfo {title} {Voltammetric determination of the stochastic
  formation rate and geometry of individual H2, N2, and O2 bubble nuclei},\
  }\href@noop {} {\bibfield  {journal} {\bibinfo  {journal} {ACS Nano}\
  }\textbf {\bibinfo {volume} {13}},\ \bibinfo {pages} {6330} (\bibinfo {year}
  {2019})}\BibitemShut {NoStop}%
\bibitem [{\citenamefont {Perez~Sirkin}\ \emph {et~al.}(2019)\citenamefont
  {Perez~Sirkin}, \citenamefont {Gadea}, \citenamefont {Scherlis},\ and\
  \citenamefont {Molinero}}]{perez2019mechanisms}%
  \BibitemOpen
  \bibfield  {author} {\bibinfo {author} {\bibfnamefont {Y.~A.}\ \bibnamefont
  {Perez~Sirkin}}, \bibinfo {author} {\bibfnamefont {E.~D.}\ \bibnamefont
  {Gadea}}, \bibinfo {author} {\bibfnamefont {D.~A.}\ \bibnamefont
  {Scherlis}},\ and\ \bibinfo {author} {\bibfnamefont {V.}~\bibnamefont
  {Molinero}},\ }\bibfield  {title} {\bibinfo {title} {Mechanisms of nucleation
  and stationary states of electrochemically generated nanobubbles},\
  }\href@noop {} {\bibfield  {journal} {\bibinfo  {journal} {J. Am. Chem.
  Soc.}\ }\textbf {\bibinfo {volume} {141}},\ \bibinfo {pages} {10801}
  (\bibinfo {year} {2019})}\BibitemShut {NoStop}%
\bibitem [{\citenamefont {Maheshwari}\ \emph {et~al.}(2020)\citenamefont
  {Maheshwari}, \citenamefont {Van~Kruijsdijk}, \citenamefont {Sanyal},\ and\
  \citenamefont {Harvey}}]{maheshwari2020nucleation}%
  \BibitemOpen
  \bibfield  {author} {\bibinfo {author} {\bibfnamefont {S.}~\bibnamefont
  {Maheshwari}}, \bibinfo {author} {\bibfnamefont {C.}~\bibnamefont
  {Van~Kruijsdijk}}, \bibinfo {author} {\bibfnamefont {S.}~\bibnamefont
  {Sanyal}},\ and\ \bibinfo {author} {\bibfnamefont {A.~D.}\ \bibnamefont
  {Harvey}},\ }\bibfield  {title} {\bibinfo {title} {Nucleation and growth of a
  nanobubble on rough surfaces},\ }\href@noop {} {\bibfield  {journal}
  {\bibinfo  {journal} {Langmuir}\ }\textbf {\bibinfo {volume} {36}},\ \bibinfo
  {pages} {4108} (\bibinfo {year} {2020})}\BibitemShut {NoStop}%
\bibitem [{\citenamefont {Ma}\ \emph {et~al.}(2021)\citenamefont {Ma},
  \citenamefont {Guo}, \citenamefont {Chen},\ and\ \citenamefont
  {Zhang}}]{ma2021dynamic}%
  \BibitemOpen
  \bibfield  {author} {\bibinfo {author} {\bibfnamefont {Y.}~\bibnamefont
  {Ma}}, \bibinfo {author} {\bibfnamefont {Z.}~\bibnamefont {Guo}}, \bibinfo
  {author} {\bibfnamefont {Q.}~\bibnamefont {Chen}},\ and\ \bibinfo {author}
  {\bibfnamefont {X.}~\bibnamefont {Zhang}},\ }\bibfield  {title} {\bibinfo
  {title} {Dynamic equilibrium model for surface nanobubbles in
  electrochemistry},\ }\href@noop {} {\bibfield  {journal} {\bibinfo  {journal}
  {Langmuir}\ }\textbf {\bibinfo {volume} {37}},\ \bibinfo {pages} {2771}
  (\bibinfo {year} {2021})}\BibitemShut {NoStop}%
\bibitem [{\citenamefont {Lohse}\ and\ \citenamefont
  {Zhang}(2015{\natexlab{a}})}]{lohse2015surface}%
  \BibitemOpen
  \bibfield  {author} {\bibinfo {author} {\bibfnamefont {D.}~\bibnamefont
  {Lohse}}\ and\ \bibinfo {author} {\bibfnamefont {X.}~\bibnamefont {Zhang}},\
  }\bibfield  {title} {\bibinfo {title} {Surface nanobubbles and
  nanodroplets},\ }\href@noop {} {\bibfield  {journal} {\bibinfo  {journal}
  {Rev. Mod. Phys}\ }\textbf {\bibinfo {volume} {87}},\ \bibinfo {pages} {981}
  (\bibinfo {year} {2015}{\natexlab{a}})}\BibitemShut {NoStop}%
\bibitem [{\citenamefont {Weijs}\ and\ \citenamefont
  {Lohse}(2013)}]{weijs2013surface}%
  \BibitemOpen
  \bibfield  {author} {\bibinfo {author} {\bibfnamefont {J.~H.}\ \bibnamefont
  {Weijs}}\ and\ \bibinfo {author} {\bibfnamefont {D.}~\bibnamefont {Lohse}},\
  }\bibfield  {title} {\bibinfo {title} {Why surface nanobubbles live for
  hours},\ }\href@noop {} {\bibfield  {journal} {\bibinfo  {journal} {Phys.
  Rev. Lett.}\ }\textbf {\bibinfo {volume} {110}},\ \bibinfo {pages} {054501}
  (\bibinfo {year} {2013})}\BibitemShut {NoStop}%
\bibitem [{\citenamefont {Lou}\ \emph {et~al.}(2000)\citenamefont {Lou},
  \citenamefont {Ouyang}, \citenamefont {Zhang}, \citenamefont {Li},
  \citenamefont {Hu}, \citenamefont {Li},\ and\ \citenamefont
  {Yang}}]{lou2000nanobubbles}%
  \BibitemOpen
  \bibfield  {author} {\bibinfo {author} {\bibfnamefont {S.-T.}\ \bibnamefont
  {Lou}}, \bibinfo {author} {\bibfnamefont {Z.-Q.}\ \bibnamefont {Ouyang}},
  \bibinfo {author} {\bibfnamefont {Y.}~\bibnamefont {Zhang}}, \bibinfo
  {author} {\bibfnamefont {X.-J.}\ \bibnamefont {Li}}, \bibinfo {author}
  {\bibfnamefont {J.}~\bibnamefont {Hu}}, \bibinfo {author} {\bibfnamefont
  {M.-Q.}\ \bibnamefont {Li}},\ and\ \bibinfo {author} {\bibfnamefont {F.-J.}\
  \bibnamefont {Yang}},\ }\bibfield  {title} {\bibinfo {title} {Nanobubbles on
  solid surface imaged by atomic force microscopy},\ }\href@noop {} {\bibfield
  {journal} {\bibinfo  {journal} {J. Vac. Sci. Technol.}\ }\textbf {\bibinfo
  {volume} {18}},\ \bibinfo {pages} {2573} (\bibinfo {year}
  {2000})}\BibitemShut {NoStop}%
\bibitem [{\citenamefont {Zhang}\ \emph {et~al.}(2006)\citenamefont {Zhang},
  \citenamefont {Maeda},\ and\ \citenamefont {Craig}}]{zhang2006physical}%
  \BibitemOpen
  \bibfield  {author} {\bibinfo {author} {\bibfnamefont {X.~H.}\ \bibnamefont
  {Zhang}}, \bibinfo {author} {\bibfnamefont {N.}~\bibnamefont {Maeda}},\ and\
  \bibinfo {author} {\bibfnamefont {V.~S.}\ \bibnamefont {Craig}},\ }\bibfield
  {title} {\bibinfo {title} {Physical properties of nanobubbles on hydrophobic
  surfaces in water and aqueous solutions},\ }\href@noop {} {\bibfield
  {journal} {\bibinfo  {journal} {Langmuir}\ }\textbf {\bibinfo {volume}
  {22}},\ \bibinfo {pages} {5025} (\bibinfo {year} {2006})}\BibitemShut
  {NoStop}%
\bibitem [{\citenamefont {Tan}\ \emph {et~al.}(2021)\citenamefont {Tan},
  \citenamefont {An},\ and\ \citenamefont {Ohl}}]{tan2021stability}%
  \BibitemOpen
  \bibfield  {author} {\bibinfo {author} {\bibfnamefont {B.~H.}\ \bibnamefont
  {Tan}}, \bibinfo {author} {\bibfnamefont {H.}~\bibnamefont {An}},\ and\
  \bibinfo {author} {\bibfnamefont {C.-D.}\ \bibnamefont {Ohl}},\ }\bibfield
  {title} {\bibinfo {title} {Stability of surface and bulk nanobubbles},\
  }\href@noop {} {\bibfield  {journal} {\bibinfo  {journal} {Curr. Opin.
  Colloid Interface Sci}\ }\textbf {\bibinfo {volume} {53}},\ \bibinfo {pages}
  {101428} (\bibinfo {year} {2021})}\BibitemShut {NoStop}%
\bibitem [{\citenamefont {Lohse}\ and\ \citenamefont
  {Zhang}(2015{\natexlab{b}})}]{lohse2015pinning}%
  \BibitemOpen
  \bibfield  {author} {\bibinfo {author} {\bibfnamefont {D.}~\bibnamefont
  {Lohse}}\ and\ \bibinfo {author} {\bibfnamefont {X.}~\bibnamefont {Zhang}},\
  }\bibfield  {title} {\bibinfo {title} {Pinning and gas oversaturation imply
  stable single surface nanobubbles},\ }\href@noop {} {\bibfield  {journal}
  {\bibinfo  {journal} {Phys. Rev. E}\ }\textbf {\bibinfo {volume} {91}},\
  \bibinfo {pages} {031003(R)} (\bibinfo {year}
  {2015}{\natexlab{b}})}\BibitemShut {NoStop}%
\bibitem [{\citenamefont {Plimpton}(1995)}]{plimpton1995fast}%
  \BibitemOpen
  \bibfield  {author} {\bibinfo {author} {\bibfnamefont {S.}~\bibnamefont
  {Plimpton}},\ }\bibfield  {title} {\bibinfo {title} {Fast parallel algorithms
  for short-range molecular dynamics},\ }\href@noop {} {\bibfield  {journal}
  {\bibinfo  {journal} {J. Comput. Phys.}\ }\textbf {\bibinfo {volume} {117}},\
  \bibinfo {pages} {1} (\bibinfo {year} {1995})}\BibitemShut {NoStop}%
\bibitem [{\citenamefont {Molinero}\ and\ \citenamefont
  {Moore}(2009)}]{molinero2009water}%
  \BibitemOpen
  \bibfield  {author} {\bibinfo {author} {\bibfnamefont {V.}~\bibnamefont
  {Molinero}}\ and\ \bibinfo {author} {\bibfnamefont {E.~B.}\ \bibnamefont
  {Moore}},\ }\bibfield  {title} {\bibinfo {title} {Water modeled as an
  intermediate element between carbon and silicon},\ }\href@noop {} {\bibfield
  {journal} {\bibinfo  {journal} {J. Phys. Chem. B}\ }\textbf {\bibinfo
  {volume} {113}},\ \bibinfo {pages} {4008} (\bibinfo {year}
  {2009})}\BibitemShut {NoStop}%
\bibitem [{\citenamefont {Saito}(1968)}]{saito1968theoretical}%
  \BibitemOpen
  \bibfield  {author} {\bibinfo {author} {\bibfnamefont {Y.}~\bibnamefont
  {Saito}},\ }\bibfield  {title} {\bibinfo {title} {A theoretical study on the
  diffusion current at the stationary electrodes of circular and narrow band
  types},\ }\href@noop {} {\bibfield  {journal} {\bibinfo  {journal} {Rev.
  Polarogr.}\ }\textbf {\bibinfo {volume} {15}},\ \bibinfo {pages} {177}
  (\bibinfo {year} {1968})}\BibitemShut {NoStop}%
\bibitem [{\citenamefont {Chen}\ \emph {et~al.}(2014)\citenamefont {Chen},
  \citenamefont {Luo}, \citenamefont {Faraji}, \citenamefont {Feldberg},\ and\
  \citenamefont {White}}]{chen2014electrochemical}%
  \BibitemOpen
  \bibfield  {author} {\bibinfo {author} {\bibfnamefont {Q.}~\bibnamefont
  {Chen}}, \bibinfo {author} {\bibfnamefont {L.}~\bibnamefont {Luo}}, \bibinfo
  {author} {\bibfnamefont {H.}~\bibnamefont {Faraji}}, \bibinfo {author}
  {\bibfnamefont {S.~W.}\ \bibnamefont {Feldberg}},\ and\ \bibinfo {author}
  {\bibfnamefont {H.~S.}\ \bibnamefont {White}},\ }\bibfield  {title} {\bibinfo
  {title} {Electrochemical measurements of single H2 nanobubble nucleation and
  stability at Pt nanoelectrodes},\ }\href@noop {} {\bibfield  {journal}
  {\bibinfo  {journal} {J. Phys. Chem. Lett}\ }\textbf {\bibinfo {volume}
  {5}},\ \bibinfo {pages} {3539} (\bibinfo {year} {2014})}\BibitemShut
  {NoStop}%
\bibitem [{\citenamefont {Popov}(2005)}]{popov2005evaporative}%
  \BibitemOpen
  \bibfield  {author} {\bibinfo {author} {\bibfnamefont {Y.~O.}\ \bibnamefont
  {Popov}},\ }\bibfield  {title} {\bibinfo {title} {Evaporative deposition
  patterns: spatial dimensions of the deposit},\ }\href@noop {} {\bibfield
  {journal} {\bibinfo  {journal} {Phys. Rev. E}\ }\textbf {\bibinfo {volume}
  {71}},\ \bibinfo {pages} {036313} (\bibinfo {year} {2005})}\BibitemShut
  {NoStop}%
\bibitem [{\citenamefont {Silbey}\ \emph {et~al.}(2022)\citenamefont {Silbey},
  \citenamefont {Alberty}, \citenamefont {Papadantonakis},\ and\ \citenamefont
  {Bawendi}}]{silbey2022physical}%
  \BibitemOpen
  \bibfield  {author} {\bibinfo {author} {\bibfnamefont {R.~J.}\ \bibnamefont
  {Silbey}}, \bibinfo {author} {\bibfnamefont {R.~A.}\ \bibnamefont {Alberty}},
  \bibinfo {author} {\bibfnamefont {G.~A.}\ \bibnamefont {Papadantonakis}},\
  and\ \bibinfo {author} {\bibfnamefont {M.~G.}\ \bibnamefont {Bawendi}},\
  }\href@noop {} {\emph {\bibinfo {title} {Physical chemistry}}}\ (\bibinfo
  {publisher} {John Wiley \& Sons},\ \bibinfo {year} {2022})\BibitemShut
  {NoStop}%
\bibitem [{\citenamefont {Van Der~Linde}\ \emph {et~al.}(2017)\citenamefont
  {Van Der~Linde}, \citenamefont {Moreno~Soto}, \citenamefont
  {Pe{\~n}as-L{\'o}pez}, \citenamefont {Rodr{\'\i}guez-Rodr{\'\i}guez},
  \citenamefont {Lohse}, \citenamefont {Gardeniers}, \citenamefont {Van
  Der~Meer},\ and\ \citenamefont {Fern{\'a}ndez~Rivas}}]{van2017electrolysis}%
  \BibitemOpen
  \bibfield  {author} {\bibinfo {author} {\bibfnamefont {P.}~\bibnamefont {Van
  Der~Linde}}, \bibinfo {author} {\bibfnamefont {{\'A}.}~\bibnamefont
  {Moreno~Soto}}, \bibinfo {author} {\bibfnamefont {P.}~\bibnamefont
  {Pe{\~n}as-L{\'o}pez}}, \bibinfo {author} {\bibfnamefont {J.}~\bibnamefont
  {Rodr{\'\i}guez-Rodr{\'\i}guez}}, \bibinfo {author} {\bibfnamefont
  {D.}~\bibnamefont {Lohse}}, \bibinfo {author} {\bibfnamefont
  {H.}~\bibnamefont {Gardeniers}}, \bibinfo {author} {\bibfnamefont
  {D.}~\bibnamefont {Van Der~Meer}},\ and\ \bibinfo {author} {\bibfnamefont
  {D.}~\bibnamefont {Fern{\'a}ndez~Rivas}},\ }\bibfield  {title} {\bibinfo
  {title} {Electrolysis-driven and pressure-controlled diffusive growth of
  successive bubbles on microstructured surfaces},\ }\href@noop {} {\bibfield
  {journal} {\bibinfo  {journal} {Langmuir}\ }\textbf {\bibinfo {volume}
  {33}},\ \bibinfo {pages} {12873} (\bibinfo {year} {2017})}\BibitemShut
  {NoStop}%
\bibitem [{\citenamefont {Cussler}(2009)}]{cussler2009diffusion}%
  \BibitemOpen
  \bibfield  {author} {\bibinfo {author} {\bibfnamefont {E.~L.}\ \bibnamefont
  {Cussler}},\ }\href@noop {} {\emph {\bibinfo {title} {Diffusion: mass
  transfer in fluid systems}}}\ (\bibinfo  {publisher} {Cambridge university
  press},\ \bibinfo {year} {2009})\BibitemShut {NoStop}%
\bibitem [{\citenamefont {Yang}\ \emph {et~al.}(2009)\citenamefont {Yang},
  \citenamefont {Tsai}, \citenamefont {Kooij}, \citenamefont {Prosperetti},
  \citenamefont {Zandvliet},\ and\ \citenamefont
  {Lohse}}]{yang2009electrolytically}%
  \BibitemOpen
  \bibfield  {author} {\bibinfo {author} {\bibfnamefont {S.}~\bibnamefont
  {Yang}}, \bibinfo {author} {\bibfnamefont {P.}~\bibnamefont {Tsai}}, \bibinfo
  {author} {\bibfnamefont {E.~S.}\ \bibnamefont {Kooij}}, \bibinfo {author}
  {\bibfnamefont {A.}~\bibnamefont {Prosperetti}}, \bibinfo {author}
  {\bibfnamefont {H.~J.}\ \bibnamefont {Zandvliet}},\ and\ \bibinfo {author}
  {\bibfnamefont {D.}~\bibnamefont {Lohse}},\ }\bibfield  {title} {\bibinfo
  {title} {Electrolytically generated nanobubbles on highly orientated
  pyrolytic graphite surfaces},\ }\href@noop {} {\bibfield  {journal} {\bibinfo
   {journal} {Langmuir}\ }\textbf {\bibinfo {volume} {25}},\ \bibinfo {pages}
  {1466} (\bibinfo {year} {2009})}\BibitemShut {NoStop}%
\bibitem [{\citenamefont {Ma}\ \emph {et~al.}(2022)\citenamefont {Ma},
  \citenamefont {Li}, \citenamefont {Pfeiffer}, \citenamefont {Eisener},
  \citenamefont {Ohl},\ and\ \citenamefont {Sun}}]{ma2022ion}%
  \BibitemOpen
  \bibfield  {author} {\bibinfo {author} {\bibfnamefont {X.}~\bibnamefont
  {Ma}}, \bibinfo {author} {\bibfnamefont {M.}~\bibnamefont {Li}}, \bibinfo
  {author} {\bibfnamefont {P.}~\bibnamefont {Pfeiffer}}, \bibinfo {author}
  {\bibfnamefont {J.}~\bibnamefont {Eisener}}, \bibinfo {author} {\bibfnamefont
  {C.-D.}\ \bibnamefont {Ohl}},\ and\ \bibinfo {author} {\bibfnamefont
  {C.}~\bibnamefont {Sun}},\ }\bibfield  {title} {\bibinfo {title} {Ion
  adsorption stabilizes bulk nanobubbles},\ }\href@noop {} {\bibfield
  {journal} {\bibinfo  {journal} {J. Colloid Interface Sci.}\ }\textbf
  {\bibinfo {volume} {606}},\ \bibinfo {pages} {1380} (\bibinfo {year}
  {2022})}\BibitemShut {NoStop}%
\end{thebibliography}
%

\end{document}